\documentclass[runningheads]{llncs}
\usepackage{graphicx}
\usepackage{url,xcolor}
\usepackage{amsmath,amssymb}
\usepackage{amsfonts,mathrsfs}

\newcommand{\nan}{\mathtt{NaN}}

\DeclareMathOperator*{\median}{median}

\usepackage{hyperref}

\begin{document}
\title{A Real-Time Region Tracking Algorithm Tailored to Endoscopic Video with Open-Source Implementation \\ \texttt{PREPRINT}}

\titlerunning{PREPRINT, under review}
\author{Jonathan P.\ Epperlein \and Sergiy Zhuk}%
\authorrunning{PREPRINT, under review}%
\institute{IBM Research Europe, Damastown, D15 HN66 Dublin, Ireland}%
\maketitle              %
\begin{abstract}
With a video data source, such as multispectral video acquired during administration of fluorescent tracers, extraction of time-resolved data typically requires the compensation of motion. While this can be done manually, which is arduous, or using off-the-shelf object tracking software, which often yields unsatisfactory performance, we present an algorithm which is simple and performant. Most importantly, we provide an open-source implementation, with an easy-to-use interface for researchers not inclined to write their own code, as well as Python modules that can be used programmatically.
\end{abstract}
\section{Introduction}
\label{sec:introduction}
\begin{figure}
	\includegraphics[width=.22\textwidth]{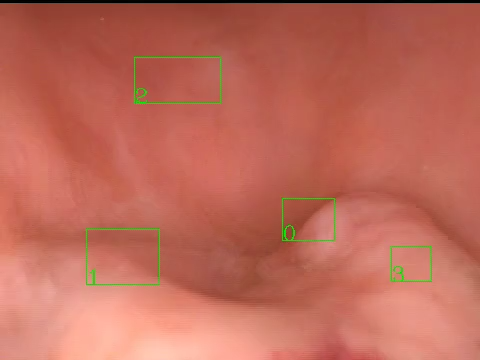}\hfill
	\includegraphics[width=.22\textwidth]{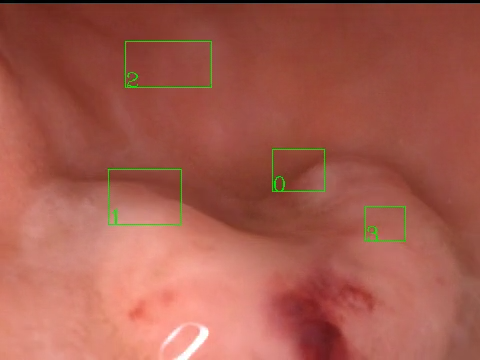}\hfill
	\includegraphics[width=.22\textwidth]{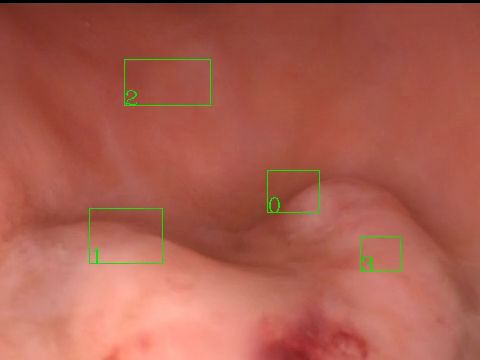}\hfill
	\includegraphics[width=.22\textwidth]{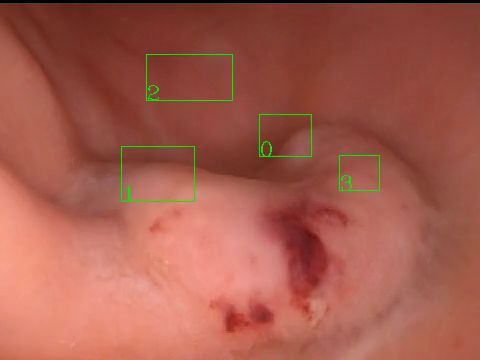}
	\caption{The proposed algorithm in action: despite considerable movement, the boxes representing regions of interest stay on the same tissue regions throughout.}
	\label{fig:intro_track}
\end{figure}
Mutltispectral imaging is now a common feature of interventional medical imaging stacks. Intraoperative use of fluorescent dyes (typically ICG) makes multispectral endoscopic video a frequently encountered source of data, as in e.g.~\cite{BJS2021,park2020artificial,khokhar2018visual}.
While in laboratory settings the specimens can be immobilized completely, in a clinical setting, imaging devices are commonly handheld and the imaged tissue is subject to constant motion, deformation, and occlusions through instruments, reflections, and fluid.
If dynamic, i.e.\ time-resolved, visual data is to be collected from different regions of interest (ROIs) of the tissue, it is thus necessary to \emph{compensate} the motion, in other words to ensure that in each consecutive frame, data is indeed collected from the same region of tissue, even as it moves to different locations within the frame, see Figure~\ref{fig:intro_track} for an illustration.
This task can be addressed manually, e.g.~\cite{khokhar2018visual}, by averaging \cite{park2020artificial}%
, or by using a more sophisticated computer vision algorithm -- the approach followed in this contribution.

Our contribution here is two-fold: We introduce a simple algorithm that can be used to correct motion in endoscopic video (or video from any source) by tracking the location of an arbitrary number or ROIs from frame to frame. We also provide an open-source implementation, which comprises Python modules that can be used programmatically by anyone inclined to do so, and a simple command-line interface (CLI) which provides access to the tracking functionality without the need to write any additional code. At the time of writing, the code, along with installation and usage notes, can be found at \url{https://github.com/IBM/optflow-region-tracker}. Algorithm and implementation have been ``field-tested'' by our collaborators and have been used for data acquisition in several prior publications \cite{BJS2021,MICCAI2020,AMIA2021,CLINIICCAI2021,Vignette2021}.

We now define the ``problem'' more rigorously, before briefly reviewing related work. Our proposed tracking algorithm is then explained in detail, before it is evaluated and compared to a suite of object trackers available in the computer vision library OpenCV~\cite{opencv_library}.%
\subsection{The ROI tracking problem}
\label{ssec:problem}
We are given a video $V$, which is really just a sequence of $T$ frames, i.e.\ $V=(I_0, I_1, \dotsc, I_{T-1})$, where each frame $I_t$ has height $h$, width $w$ and either 1 or 3 channels. The ROIs $R_0=(r_0^1, r_0^2, \dotsc, r_0^{N})$ are specified by their positions in the initial frame $I_0$, and each ROI $r^i_0$ is an axis-parallel rectangle $r^i_0 = (x^i_0, y^i_0, w^i_0, h^i_0)$. Note that by convention, the axis origin is the left-\emph{top} corner of the frame, with the $y$-axis pointing \emph{down}, so that the point $(x^i_0, y^i_0)$ is the left-top corner of the $i$-th specified ROI, and $w^i_0$, resp.\ $h^i_0$ are its width, resp.\ height.
We are now interested in estimating the subsequent ROI positions $(R_1, \dotsc, R_{T-1})$ so that whatever was initially inside the box defined by $r^i_0$ (such as a specific region of tissue) is inside $r^i_t$ in frame $I_t$.

It should be clear that whatever starts out as an axis-parallel rectangle will not remain one: rotation and perspective changes alone transform each rectangle to a general quadrilateral, and deformations  warp straight lines into general curves. A complete agreement between predicted location $r^i_t$ and the ROI's true contents $r^i_{t,\text{true}}$ is hence not achievable. As a measure of agreement between the rectangle $r^i_{t}$ and the general shape $r^i_{t,\text{true}}$ we adopt the (rasterized) Jaccard index (also called Intersection-over-Union):
\begin{equation}
	J(r^i_{t,\text{true}},r^i_{t}) = \frac{\text{number of pixels inside } r^i_{t,\text{true}} \text{ \emph{AND} } r^i_t}
										{\text{number of pixels inside } r^i_{t,\text{true}} \text{ \emph{OR} } r^i_t}.																		\label{eq:jaccard}
\end{equation}
A Jaccard index of 1 thus indicates perfect agreement, whereas 0 corresponds to completely disjoint sets.

We assume that all ROIs are visible in all frames, or we accept that if a ROI $i$ leaves the view at some point $t$, all subsequent locations $r^i_{t+1}, r^i_{t+2}, \dotsc$ will be marked as $\nan$ (not a number), i.e.\ tracking for this ROI ends at $t$ even if it enters the view again later, it can not be reacquired.

\subsection{Related Work}
\label{ssec:related}
While this at first appears to be a classical object-tracking problem, it turns out that many object trackers perform quite poorly, see Section~\ref{sec:results}. However on further reflection, this is not surprising: the typical object-tracking task is concerned with physical objects moving freely and independently in front of a largely static background. For example, the popular KITTI dataset~\cite{KITTI} contains street scenes with moving cars and pedestrians, and in a clinical context, the Cholec80 dataset~\cite{Cholec80} tracks the locations of surgical instruments. Hence, algorithms typically attempt to build a model of the object, e.g.~\cite{Boosting} uses a mix of Haar-like features, orientation histograms, and local binary patterns, to describe the object and build a classifier that can distinguish it from other regions in the same video. 

In contrast to that, the ROIs under consideration here often have very little ``objectness'' about them, in the sense that there is little distinguishing them from the rest of the video frame, there often are no edges or other salient points; however a ROI's movement is \emph{not} independent of the surrounding image, quite the opposite, typically the entire view will be filled by connected tissue (with the exception of occasional surgical tools, reflections, and occlusion by e.g.\ blood or water). One could thus say that the problem addressed here is outside the design specification of an off-the-shelf object tracking algorithm, that some of them still perform as well as they do in Section~\ref{sec:results} is thus a testament to their robustness and generalizability.

Approaches specific to endoscopic video can also be found in the literature, e.g.~\cite{Zhan2020} uses a combination of ORB key points and the MedFlow tracking algorithm, and \cite{ye2016online} describes an multi-step algorithm solving not only the tracking problem, but also able to reacquire a ROI if it reappears. However, to the best of our knowledge, no implementations of the described algorithms are available, and so they are not accessible to the average medical researcher; we can also not include them in our comparisons.

\section{Tracking Algorithm Based on Optical Flow}
\label{sec:alg}
Our proposed region tracking algorithm obtains a \emph{global} estimate of the image deformation, and from that computes a \emph{local} transformation for each ROI to find its new location. Specifically, for each pair of frames $I_t, I_{t+1}$, the optical flow field $U_t(x,y)$ is estimated, and subsequently for each ROI $r^i_t$, a local flow field, constrained to preserve the ROI as an axis-parallel rectangle, is computed to move the ROI to its new location $r^i_{t+1}$. 

\subsection{Optical Flow}
\label{ssec:optflow}
The concept of optical flow is a very basic one in computer vision, hence the available literature is vast, and the present section is just a rudimentary primer. For more background, consider~\cite{zimmer2011optic,DISflow} and the references at \cite{middlebury}.

Very roughly, the \emph{optical flow} $U_t(x,y)=(u_t(x,y), v_t(x,y))$ is a 2-D flow field, i.e.\ it assigns to each pixel $(x,y)$ a flow $u_t(x,y)$ in $x$-direction, and a flow $v_t(x,y)$ in $y$-direction, which (approximately) warps the frame $I_{t}$ into the next frame $I_{t+1}$ in the sense that 
\[
	I_{t}\bigl( x+u_t(x,y), y+v_t(x,y) \bigr) \approx I_{t+1}(x,y);
\]
in words: the color of pixel $(x,y)$ in frame $I_t$ moved to location $(x+u_t(x,y), y+v_t(x,y))$ in frame $I_{t+1}$.

The two main practical issues are that $I_{t+1}$ is never a mere rearrangement of the pixels in $I_{t}$, and that there is no unique ``best'' flow field $U_t$\footnote{to see that, consider the case where two pixels in $I_t$ have the same color as some pixel $(x,y)$ in $I_{t+1}$ -- which of them should flow to $(x,y)$?}. The latter point is addressed by the reasonable assumption that a pixel's movement is not drastically different from nearby pixels', which means the flow field may be assumed to be smooth. Hence, most optical flow algorithms minimize a mixture of two penalty terms, the ``data term'' $C_D$ measuring the disagreement between the warped frame $I_t$ and the next frame $I_{t+1}$, and the ``smoothness term'' $C_S$, quantifying the smoothness of the flow field:
\[
	U_t = \arg\min_{(u,v)} C_D\Bigl(I_{t}\bigl( x+u(x,y), y+v(x,y) \bigr) - I_{t+1}(x,y)\Bigr) + \lambda C_S\bigl( (u, v) \bigr),
\]
the weight $\lambda>0$ controls the relative importance of the smoothness. It is the smoothness term that makes the optical flow a global estimate: in regions of low or no texture, the data term yields no constraints on the flow, but the smoothness term will force the flow estimate to be a smooth interpolation of the flows computed in surrounding areas with more texture.

There are 100s of algorithms to compute the optical flow between two frames; we are using DISflow~\cite{DISflow} because it is the fastest and most precise of the ones available in OpenCV. However, what follows is completely independent of the specific algorithm used, as long as it computes a flow field $U$.
\subsection{Region Tracking}
Given the flow $U_t$, it now can be used to move the pixels in ROI $r^i_t$ to estimate the next location $r^i_{t+1}$. Simply applying the flow at the corners or along the edges of the ROI will warp it from its rectangular shape, and discard all the information from the ROI's interior. Instead, we propose to aggregate the flow inside a given ROI in a way that preserves its shape. 

The simplest such aggregation yields a constant flow vector $\bar{U}^i_t = (\bar{u}^i_t, \bar{v}^i_t)$ for each ROI, namely the median of $u_t$ and $v_t$ inside $r^i_t$. It is computed as:
\begin{align*}
	\bar{u}^i_t & = \median \{ u_t(x,y) \;|\; x^i_t\leq x \leq x^i_t+w^i_t, y^i_t\leq y \leq y^i_t+h^i_t\} \\
	\bar{v}^i_t & = \median \{ v_t(x,y) \;|\; x^i_t\leq x \leq x^i_t+w^i_t, y^i_t\leq y \leq y^i_t+h^i_t\}.
\end{align*}
The estimated next location is then
\begin{equation}
	r_{t+1}^i = (x_t^i + \bar{u}^i_t, y^i_t + \bar{v}^i_t, w^i_t, h^i_t). \label{eq:update:median}
\end{equation}
Due to the use of the median (instead of the mean), this is robust to outliers in the flow $U_t$. It preserves the shape of the ROI exactly, i.e.\ the height and width never change, and performs remarkably well. This version of our tracker is conceptually very similar to the MedFlow tracker~\cite{MedFlow}, with the difference that we do not compute key points, and scale changes are not estimated here.

To accommodate changes in scale, which for instance occur if the camera is moved closer to, or farther from, the tissue, we provide a second aggregation method, which allows scaling of ROIs, while preserving their shape as axes-parallel rectangles. It is straightforward to see that a flow field preserving axis-parallel lines has to satisfy $u(x,y) = u(x)$ and $v(x,y)=v(y)$, i.e.\ the flow in $x$-direction has to be independent of the $y$-coordinate, and vice versa. The simplest function satisfying this constraint is an affine function $u(x,y) = \tau_x + \sigma_x x$, $v(x,y) = \tau_y + \sigma_y y$. This has a straightforward interpretation: warping by such a flow corresponds to translation by $(\tau_x, \tau_y)$ and anisotropic scaling along the $x$ and $y$ axes. 

Specifically, for each ROI $r^i_t$, we compute scaling and translation by fitting an affine function to the flow field within (dependence on $t$ omitted for better readability):
\begin{align*}
	(\tau_x^i, \sigma_x^i) &= \arg\min_{(p,s)} \sum_{x=x^i}^{x^i+w^i}\sum_{y=y^i}^{y^i+h^i} \bigl| u_t(x,y) - (p + sx) \bigr|^2 \\
	(\tau_y^i, \sigma_y^i) &= \arg\min_{(p,s)} \sum_{x=x^i}^{x^i+w^i}\sum_{y=y^i}^{y^i+h^i} \bigl| v_t(x,y) - (p + sy) \bigr|^2 
\end{align*}
and the new location is
\begin{equation}
		r_{t+1}^i = \bigl(x_t^i + \sigma_x^i x_t^i +\tau_x^i,\; y^i_t + \sigma_y^i y_t^i  +\tau_y^i,\; (1+\sigma_x^i)w^i_t,\; (1+\sigma_y^i)h^i_t \bigr),  \label{eq:update:affine}	
\end{equation}
hence the ROI is scaled along the axes by $(1+\sigma_x)$, resp.\ $(1+\sigma_y)$ (with respect to the origin, which is the left-top corner of the frame) and then translated by $(\tau_x^i, \tau_y^i)$.

\section{Evaluation Results}
\label{sec:results}
The proposed algorithm is evaluated using videos generated in the following way. From each of 17 liver laparascopies and 37 colonoscopies, 2 frames at different time points were extracted, for a total of 108 initial frames of resolution 480 $\times$ 360 each. In each frame, 10 ROIs were selected at random.
To simulate motion, a sequence of 50 random projective transforms were then applied to each of the initial frames, and at the same time to each of the ROIs, yielding a synthetic video along with the ground truth locations $(r^1_0, \dotsc, r^{10}_0),\dotsc, (r^1_{50}, \dotsc, r^{10}_{50})$ for each of the 10 ROIs. 
A projective transform consists of translation, rotation, scaling, shearing, and elation, and is the most general transformation that preserves straight lines, see~\cite{szeliski2010computer} for details. Note that it follows that the ground truth locations are not rectangular anymore, but only general quadrilaterals, but their agreement with the rectangular estimates from the tracking algorithms can be easily evaluated as in~\eqref{eq:jaccard}.

Additionally, saturated regions of random sizes were added to simulate reflections, which are commonly encountered in endoscopic applications, before supplying each new frame to the trackers.

To keep the videos realistic, shear (3 pixels), scaling (between 0.95 and 1.05), and translation velocity (3 pixels/frame) were limited to be small; the random rotations were bounded by either 0\textdegree, $\pm5$\textdegree, or $\pm10$\textdegree, and the number of added reflections was either 0, 10, or 25. This process yielded a total of $108\cdot50\cdot3\cdot3=48600$ video frames.

We evaluated our tracker with both, the median aggregation as in~\eqref{eq:update:median}, and the affine aggregation as in~\eqref{eq:update:affine}, alongside the OpenCV~\cite{opencv_library} implementations of KCF~\cite{KCF}, MedianFlow~\cite{MedFlow}, the boosting-based tracker~\cite{Boosting}, CSRT~\cite{CSRT}, MIL~\cite{MIL}, and TLD~\cite{TLD}. Since Boosting, CSRT, MIL and TLD are an order of magnitude slower than the other trackers, they were evaluated on the first 10 frames of each video only, to keep computation times manageable. All algorithms' parameters are set to their defaults.

\begin{figure}
	\centering
	\includegraphics[width=.9\textwidth]{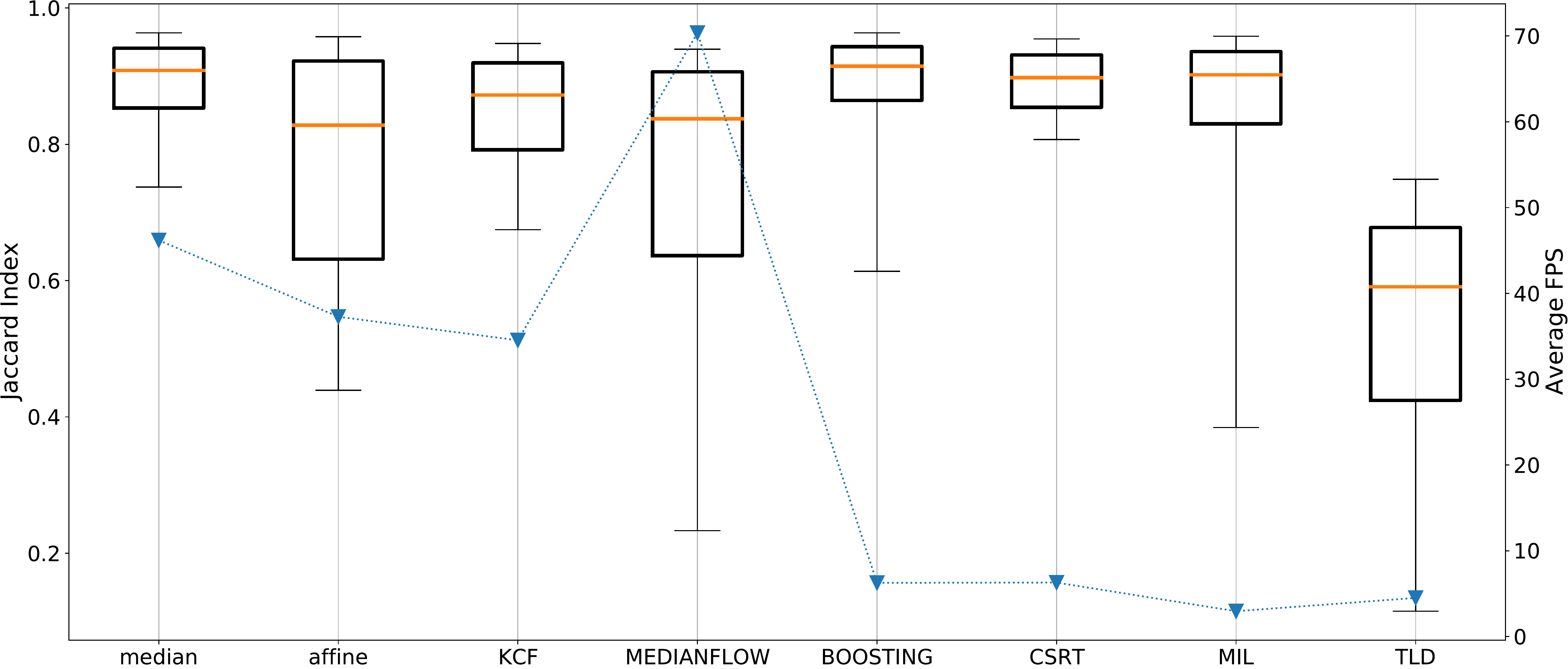}
	\caption{Performance of all trackers over all frames and ROIs. The boxes illustrate the distribution of the Jaccard indices, while the triangles indicate the number of frames processed in 1 second; clearly, Boosting, CSRT, MIL and TLD are significantly slower.}%
	\label{fig:boxplot_all}
	\centering
	\includegraphics[width=.9\textwidth]{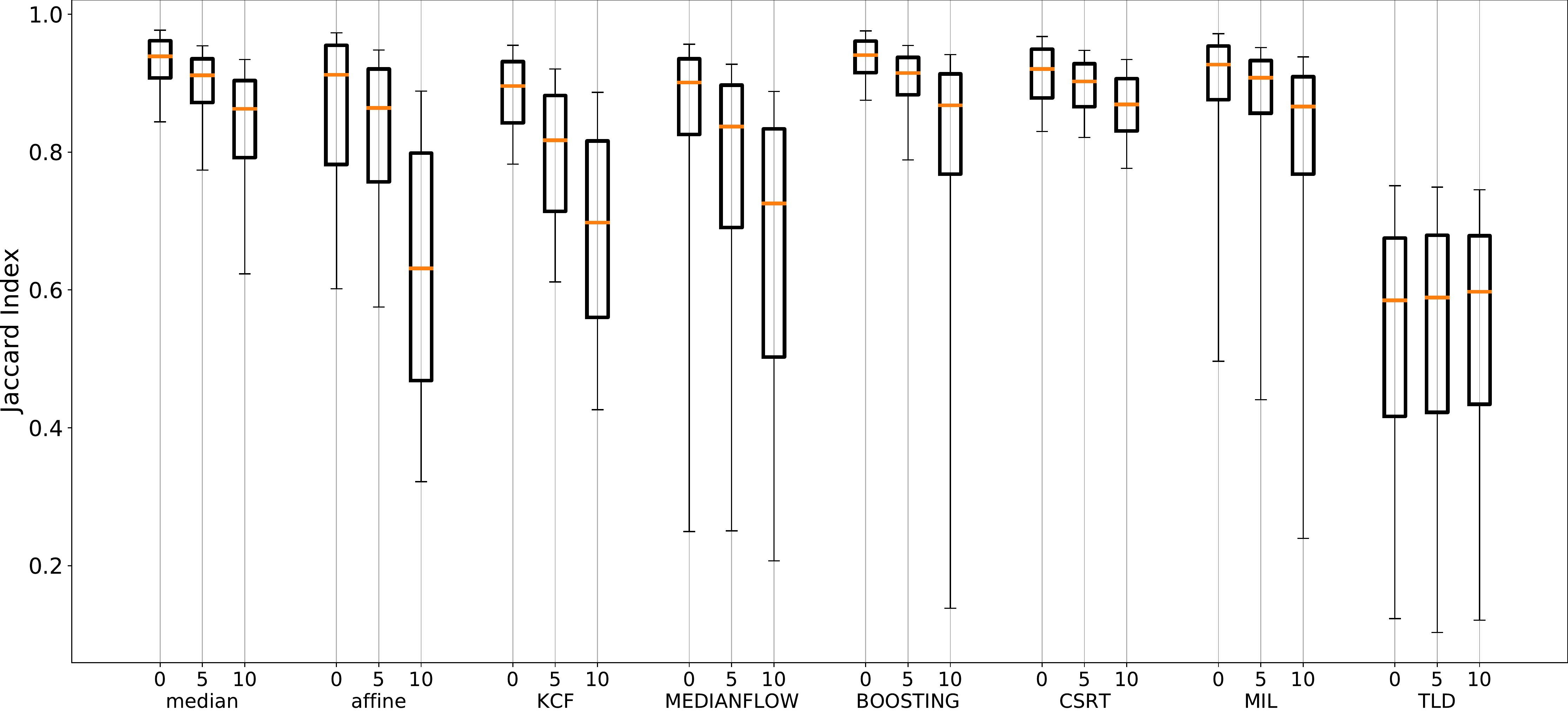}
	\caption{Figure~\ref{fig:boxplot_all}, with the influence of rotation considered separately. The 3 bars for each tracker are computed from its performance on frames generated with no rotation, rotations of less than 5\textdegree, and rotations of less than 10\textdegree, from left to right.}
	\label{fig:boxplot_rot}
	\centering
	\includegraphics[width=.9\textwidth]{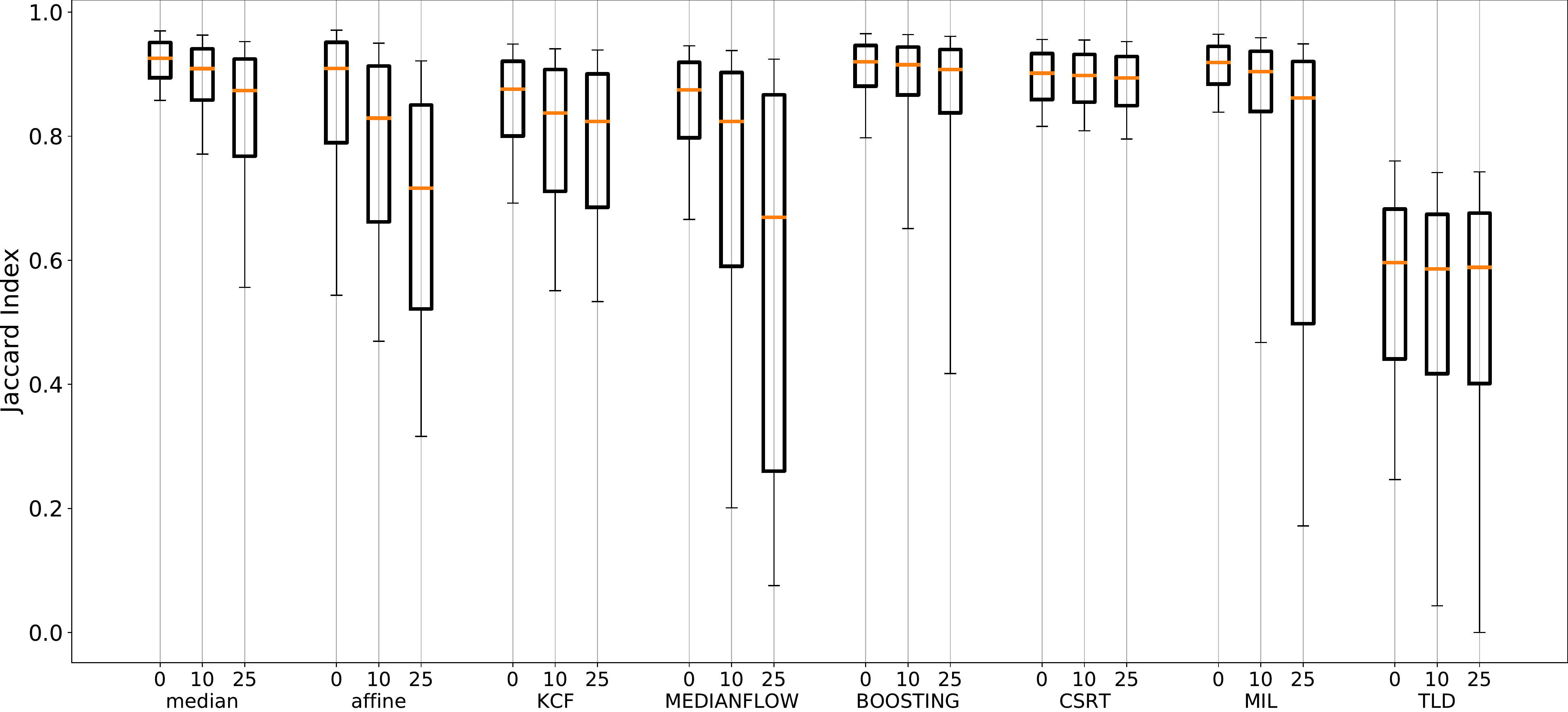}
	\caption{Figure~\ref{fig:boxplot_all}, with the influence of reflections considered separately. The 3 bars for each tracker are computed from its performance on frames generated with no reflections, 10 reflections, and 25 reflections per frame, from left to right.}
	\label{fig:boxplot_ref}
\end{figure}

All experiments were run on a MacBook Pro with a 2.6 GHz 6-Core Intel Core i7 CPU and 16GB of RAM, using Python 3.6 and OpenCV 3.4.7. The results are illustrated in Figures~\ref{fig:boxplot_all}--\ref{fig:boxplot_ref}.
\subsection{Discussion}
The proposed tracker with median flow is the clear winner among the trackers with real-time performance, maintaining a Jaccard index of 0.85 for 75\% of all frames and ROIs; however considering the effect of rotations and reflections separately also indicates its limitations, as performance degrades considerably with the addition of each. In fact, all trackers struggle with rotations and reflections, except CSRT, which remains relatively unaffected and maintains its high performance, albeit at the cost of roughly 10-fold decrease in FPS.

Based on this test dataset, there is no case to be made for the affine flow aggregation as given in~\eqref{eq:update:affine}, however in our experience there have been videos for which the affine tracker was able to track ROIs when the median-based tracker failed. We intend to investigate this further in future research.

\subsection{A Possible Use Case}
As a typical use case consider~\cite{son2019quantitative}, where the colon is imaged using a multispectral endoscope during and after ICG administration. As a result of the ICG perfusing, the infrared image lights up as ICG arrives in the tissue and fluoresces, and goes dark again as the ICG is washed out. It is conjectured that the shape of the intensity-vs-time curve of different regions of the colon can be used to quantify the quality of their perfusion. 
 
The provided tool can be used to collect these curves for an arbitrary number of regions:\\
\verb! >> python region_tracking.py --separate_vids --show-tracking! \\
The \verb!--separate_vids! option means that the user will be prompted for two video files: the first containing the visible light image, which will be used for tracking, and the second the infrared image, which will be used for data collection. Options for cases where the visible light and infrared views are separate panels in a merged video are also provided. A small Graphical User Interface (GUI) will open, in which the user can select the ROIs to be tracked. Tracking then commences until stop, and upon completion, the collected data is written into CSV files, ready for further processing. See the Supplementary Material for a screen cap video of the tool in action.

\section{Conclusion and Caveats}
\label{sec:concl}
While the algorithm and its implementation performed well in the evaluation of Section~\ref{sec:results}, and has performed well for us in practice, a few drawbacks need to be emphasized.
\begin{itemize}
 \item Unlike the other evaluated trackers, our tracker currently does not supervise its own performance, i.e.\ it does not assess if the region within the estimated ROI location is visually similar to the ROI from frame 0, or the previous frame. This is in active development, and a forward-backward error similar to the one proposed in~\cite{MedFlow} is being implemented.
 \item The tracker is not able to re-acquire a ROI if it was lost from view at any time; it is also not able to recover when its position has drifted. As a result, its performance degrades over time, and tracking for longer than $\sim\!15$ minutes requires a steady hand by the imaging surgeon.
 \item The tracker is robust to quality issues such as reflections, occlusions, motion blur, compression artefacts, and interlacing, only as much as the underlying optical flow estimation algorithm is. The implementations in OpenCV offer no option to specify a mask of ignored pixels, so poor-quality video can cause problems.
\end{itemize}
As in most applications to real data, your mileage may vary, and we suggest to not trust the collected data blindly, but to manually supervise the tracker's performance: a video with the estimated ROI locations is generated automatically and allows for a quick quality check.

\end{document}